\documentclass[%
 reprint,
 showpacs,preprintnumbers,
 amsmath,amssymb,
 aps,
prb,
]{revtex4-1}
\usepackage{dcolumn}
\usepackage[dvipdfm]{graphicx}
\usepackage{subfigure,color}
\usepackage{mathrsfs}

\makeatletter
\def\btt#1{\texttt{\@backslashchar#1}}
\DeclareRobustCommand\bblash{\btt{\@backslashchar}} \makeatother

\begin{document}

\title{Purely Surface Electric Transport Phenomenon on a Three-Dimensional Topological Insulator}

\author{Tetsuro Habe$^1$ and Yasuhiro Asano$^{1,2}$}
\affiliation{$^1$Department of Applied Physics,
Hokkaido University, Sapporo 060-8628, Japan}
\affiliation{$^2$Center for Topological Science \& Technology,
Hokkaido University, Sapporo 060-8628, Japan}

\date{\today}

\begin{abstract}
We theoretically study the electric current induced by the temporally oscillating magnetic field onto a three-dimensional topological insulator based on the linear response theory. 
The electric current flows on its surface because of the unique characters of the surface states: the linear dispersion in two-dimension and the chiral spin texture on the Fermi surface.
In the bulk region, on the other hand, the current is absent because of the degeneracy for the bulk bands.
The response function has a peak structure when the frequency of the magnetic field is equal to twice of the Fermi energy of the surface states measured from the Dirac node. This feature is unique to the surface states on the three-dimensional topological insulator.
\end{abstract}


\maketitle
\section{Introduction}
The presence of two-dimensional metallic surface states on three-dimensional(3D) topological insulators(TIs) is the sign of the topologically non-trivial bulk insulating states\cite{Fu2007,Fu2007-2,Moore2007,Chen2009}.
Similar to the massless fermions in particle physics, the surface metallic states have the two characteristic features: the linear energy dispersion and the chiral spin texture on the Fermi surface\cite{Hsieh2009,Hsieh2009-2,Roushan2009}.
Thus 3D TIs provide a testing ground for examining electric properties of such fundamental particles.

A number of theoretical studies have suggested unusual transport phenomena on the surface of 3D TIs\cite{Burkov2010,Wang2011,Adam2012,Li2012}.
In real materials, however, the measured electric current under the bias voltage includes two contributions:
the current flowing on the surface and that flowing through the bulk.
Thus 
the high resistive bulk samples\cite{Ren2010,Ando2013}, the thin films\cite{Lu2010,Steinberg2010,Checkelsky2011}, and the natural heterostructure of Bi$_2$Se$_3$ and Pb$_2$Se$_3$\cite{Nakayama2012} are necessary to study the transport phenomena on the surface by eliminating the contribution from the bulk.
Even so, the contribution of the bulk states to the electric current may be unavoidable because the bulk insulating gap in real materials would be smaller than that in the ideal sample due to uncontrollable mid-gap levels and unscreened electric potential by doping\cite{Skinner2012}.
It is still very difficult to selectively measure the current on the surface.
In such situation, we need theoretical proposals for experimental setup which enable to detect pure surface current.
This is an important open issue for understanding the basic physics of Dirac particle and for considering the application of it.

In this paper, we study the electric current on the surface of 3D TIs induced by a temporally oscillating magnetic field as shown in Fig. 1.
The Zeeman coupling to the magnetic field causes the spin polarization which generates the net current of the surface states.
This is because the directions of the spin and those of the velocity are locked to each other in the surface states.
We analytically calculate the electric current on the basis of the linear response theory.
The electric current is sensitive to the ratio of $\omega/\mu_F$, where $\omega$ and $\mu_F$ are the frequency of oscillating magnetic field and the Fermi energy with respect to the Dirac point, respectively. Because of the chiral spin texture on the Dirac cone, the current has a peak at $\omega/2\mu_F=1$, which is an unique transport phenomenon to the Dirac particle.
In contrast to the current under the bias voltage, the current induced by the oscillating magnetic field flows only through the surface.
The electric current in the bulk is absent because of the spin degeneracy at each momentum in the bulk states.
We also show that these are unique features of 3D TIs and are not observed in other materials having the Dirac cones such as graphene\cite{Neto2009} and silicene\cite{}.

\begin{figure}[htbp]
 \includegraphics[width=60mm]{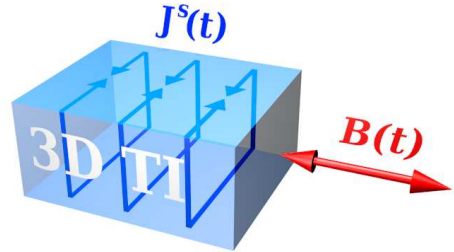}
\caption{The three-dimensional topological insulator in oscillating magnetic field.
The current flowing through the surface is perpendicular to the magnetic field.
 }\label{fig1}
\end{figure}

\section{Electromagnetic Induction on Surface}
\begin{figure}[htbp]
 \includegraphics[width=40mm]{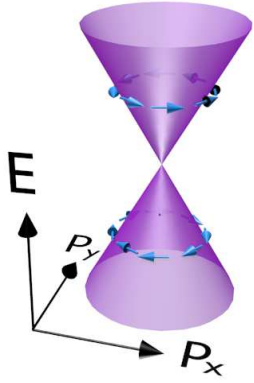}
\caption{The energy dispersion of the surface states.
The surface states have the chiral spin texture illustrated by the arrows.
 }\label{fig2}
\end{figure}

We consider the two-dimensional electric states on the $xy$ surface of the 3D TIs.
The surface states are described by the $2\times2$ Dirac Hamiltonian,
\begin{align}
H_0=v(p_x\sigma^y-p_y\sigma^x)-\mu_F,
\end{align}
with the momentum $p_\nu$ for $\nu=x,y$, the Pauli matrix $\sigma^\nu$ for $\nu=x,y$ and $z$, the Fermi velocity $v$, and the Fermi energy $\mu_F$.
The energies of the surface states form a linear dispersion $\varepsilon_{\boldsymbol{p},\pm}=\pm vp-\mu_F$ so called Dirac cones.
In the two-dimensional Brillouin zone on a single surface, two Dirac cones touch at $\boldsymbol{p}=0$ shown in Fig. \ref{fig2}.  
At a single momentum, there are two eigenstates $|\boldsymbol{p},+\rangle$ and $|\boldsymbol{p},-\rangle$ 
belonging to the upper and lower cones, respectively.
The two eigenstates have opposite spin,
\begin{align}
\langle \boldsymbol{r}|\boldsymbol{p},\pm\rangle=\frac{1}{\sqrt{2}}\begin{pmatrix}
 i\\
\pm e^{i\theta_{\boldsymbol{p}}}
\end{pmatrix}e^{i\boldsymbol{p}\cdot\boldsymbol{r}},\label{wavefunction}
\end{align}
with $p_x/p_y=\tan\theta_{\boldsymbol{p}}$.

We calculate the electric current induced by a temporally oscillating magnetic field on the surface of 3D TIs as shown in Fig.\ref{fig1}.
The magnetic field $\boldsymbol{B}(t)=(B_xe^{-i\omega t},0,0)$ couples to the orbital and the spin of the surface states.
The effects are represented by
\begin{align}
H'=\boldsymbol{j}^S\cdot\boldsymbol{A}(\boldsymbol{r},t)+g\mu_B\boldsymbol{B}(t)\cdot\boldsymbol{\sigma},\label{input}
\end{align} 
with the Lande g-factor $g$, the Bohr magneton $\mu_B$ and the electric current $\boldsymbol{j}^S=e{\boldsymbol{v}}$ on the surface.
Here $\boldsymbol{A}(\boldsymbol{r},t)=(0,0,yB_xe^{-i\omega t})$ is a gauge vector with $\nabla\times\boldsymbol{A}=\boldsymbol{B}$.
Because of the spin-momentum locking in the surface states, 
the velocity operator can be written by using the spin operator
\begin{align}
{\boldsymbol{v}}=\nabla_{\boldsymbol{p}} H_0
=v(\sigma^y,-\sigma^x,0),\label{velocity}
\end{align} 
with the Fermi velocity $v$.
Thus the electric current operator is proportional to the spin operator as
\begin{align}
{\boldsymbol{j}}^S=ev(\sigma^y,-\sigma^x,0),\label{current}
\end{align}
with the electric charge $e$.
According to the relation between the electric current and the spin, the magnetic field in Eq. (\ref{input}) couples to the electric current through the Zeeman coupling.
For better understanding the coupling, we rewrite the Zeeman term by using the effective field $E_B(t)$ as
\begin{align}
g\mu_B\boldsymbol{B}\cdot\boldsymbol{\sigma}=j_y^S\frac{E_B(t)}{\omega},\label{EffE}
\end{align}
with $E_B(t)=g\mu_B\omega B_x\exp[-i\omega t]/e$.
The Zeeman term is equivalent to the coupling between the current and the effective electric field.
This is a characteristic feature of the surface states on 3DTIs.
Thus, for nonzero frequency, the Zeeman coupling induce the electric current $j^S$.
The magnetic fields induce not only the magnetization but also the electric current
because the accumulation of the spin is equal to the non-zero net current on the surface.
In the linear response theory, the current induced by the magnetic field is calculated by
\begin{align}
\langle j^S\rangle_\nu(t)=-i\int^t_{-\infty}dt'\mathrm{Tr}\left\{\rho_0\left[\hat{j}_\nu^S(t),\hat{H}'(t')\right]\right\},\label{KuboF}
\end{align}
where we use the density matrix $\rho_0=\mathrm{Tr}[e^{\beta(-K_0+\Omega)}]$ with temperature $\beta^{-1}$, the total Hamiltonian of the all particles $K_0$ and the thermodynamic potential $\Omega$. 
The time-dependent field operator is defined by 
\begin{align*}
\hat{O}(t)=\sum_{m,n}\hat{\psi}_m^\dagger(t) O\hat{\psi}_n(t)
\end{align*}
 with the annihilation operator $\hat{\psi}_m(t)=e^{iK_0t} \psi_m e^{-iK_0t}$ on the eigenstate $m$ by using the unit of $\hbar=1$.
In the presence of the inversion symmetry, the first term in Eq.~(\ref{input}) vanishes because the electric current $j_\nu^S$ and vector potential $\boldsymbol{A}(\boldsymbol{r},t)$ are antisymmetric and symmetric under the spatial inversion of $\nu\rightarrow -\nu$, respectively.
As a result, the left-hand side of Eq. (\ref{KuboF}) is antisymmetric under the spatial inversion $\nu\rightarrow-\nu$, whereas the right-hand side of Eq. (\ref{KuboF}) is symmetric.
Even in the presence of the non-correlated disorder, the contribution from the first term in Eq. (\ref{input}) is negligible
because the current in Eq. (\ref{KuboF}) is the averaged value on the surface.
Therefore we consider the contribution from only the Zeeman term in what follows.

The response function is defined by 
\begin{align*}
\langle j\rangle_\nu(t)=\mathcal{R}_\nu(\omega)B(t),\\
\mathcal{R}_\nu(\omega)=R_\nu(\omega)+i\tilde{R}_\nu(\omega).
\end{align*}
In the absence of the disorder, the real part of the response function is represented by
\begin{align*}
R_\mu(\omega)=&-\frac{g\mu_B}{eV}e^{\beta\Omega}\sum_{m,n} (e^{-\beta E_n}-e^{-\beta E_m})\\
&\times \langle n|j_\mu^S|m\rangle\langle m| j_\nu^S|n\rangle\mathcal{P}\frac{1}{\omega+ \varepsilon_n -\varepsilon_m}
\end{align*}
with the area of the surface $V$. The symbol $\mathcal{P}$ is the principle value of the following function.
Here $E_n$ and $|n\rangle$ represent the total energy and the eigenstate of all particles, respectively.
The single particle energy is denoted by $\varepsilon_n$.
In the clean limit, the initial state $|n\rangle$ and the intermediate state $|m\rangle$ must have the same momentum. 
Thus the matrix element $\langle n|j_\mu^S|m\rangle$ is non-zero under the condition of
\begin{align}
\varepsilon_m+\mu_F=-(\varepsilon_n+\mu_F)=vp.
\end{align}
Using Eq. (\ref{wavefunction}), the response function $R_\mu(\omega)$ is calculated by integrals for the angle $\theta_{\boldsymbol{p}}$ and the magnitude $p$ of the momentum, as described in Appendix,
\begin{align}
R_\mu(\omega)
=&-\frac{g\mu_B}{e}\int_0^{2\pi} d\theta_{\boldsymbol{p}}\langle \boldsymbol{p},+|{{j}_\mu^S}|\boldsymbol{p},-\rangle\langle \boldsymbol{p},-|{{j}_y^S}|\boldsymbol{p},+\rangle\nonumber\\
&\times\int^{p_g}_{0} dp\; \left\{f_F(\beta,vp-\mu_F)-f_F(\beta,-vp-\mu_F)\right\}\nonumber\\
&\times
\left(\mathcal{P}\frac{p}{\omega+ 2vp}-\mathcal{P}\frac{p}{\omega- 2vp}\right),\label{int}
\end{align}
with the bulk gap energy $vp_g$ and the Fermi distribution function $f_F(\beta,\varepsilon)=(1+e^{\beta \varepsilon})^{-1}$.
Since the current density operator is proportional to the spin operator $\sigma^\nu$, 
the integral for $\theta_{\boldsymbol{p}}$ is
\begin{align*}
&\int_0^{2\pi} d\theta_{\boldsymbol{p}}\langle \boldsymbol{p},+|{{j}_\mu^S}|\boldsymbol{p},-\rangle\langle \boldsymbol{p},-|{{j}_y^S}|\boldsymbol{p},+\rangle\\
=&e^2v^2\int_0^{2\pi} d\theta_{\boldsymbol{p}}\epsilon_{\mu\nu z}P^\nu P^x,
\end{align*}
where the transition probability $P^\nu=\langle \boldsymbol{p},+|\sigma^\nu|\boldsymbol{p},-\rangle$ is 
\begin{align}
P^x=i\cos\theta_{\boldsymbol{p}},\;\;P^y=i\sin\theta_{\boldsymbol{p}}.
\end{align} 
Using $P^\mu$, the current along $y$ axis remains, whereas the current along $x$ axis vanishes 
because of
\begin{align*}
\int_0^{2\pi} d\theta_{\boldsymbol{p}}\langle \boldsymbol{p},+|{{j}_\mu^S}|\boldsymbol{p},-\rangle\langle \boldsymbol{p},-|{{j}_y^S}|\boldsymbol{p},+\rangle
=&\begin{cases}
\pi e^2v^2&(\mu=y)\\
0&(\mu=x)
\end{cases}.
\end{align*}
Therefore the electric current $\langle j^S\rangle_y$ perpendicular to the magnetic field remains on the surface. 

In the low-temperature limit $\beta\rightarrow\infty$, the Fermi distribution function is equal to the step function,
\begin{align*}
\lim_{\beta\rightarrow\infty}f_F(\beta,\varepsilon)=\theta(-\varepsilon).
\end{align*}
In this case, the step function restricts the range of integral to $[\mu_F,\;vp_g]$. The integral for $p$ is easily calculated as shown in Appendix,
\begin{align*}
R_\mu(\omega)=&-\frac{\pi eg\mu_B}{2}
\left\{2(\mu_F-\xi_g)\right.\\
&\left.+\frac{\omega}{2}\ln\frac{\xi_g-\omega/2}{\xi_g+\omega/2}
 \frac{|\omega/2-\mu_F|}{|\mu_F+\omega/2|}\right\},
\end{align*}
with $\xi=vp$ and $\xi_g=vp_g$.
The first term is the magnetization current and is not observed in experiments.
The second term, on the other hand, represents the dissipative and observable current which vanishes at $\omega=0$.
The observable electric current reaches a peak at $\omega=2\mu_F$ as
\begin{align}
R_\mu(\omega)\simeq&\frac{\pi eg\mu_B\omega}{4}\ln
 \frac{\omega}{|\omega/2-\mu_F|}.
\end{align}
At $\omega=2\mu_F$, the oscillating magnetic field can excite an electron with energy $-\mu_F$ in the lower cone to just the Fermi level without changing its momentum.
In this process, the conducting electron is provided on the Fermi surface.
Therefore the response function takes its maximum at $\omega=2\mu_F$.
The current does not vanish even in the presence of potential disorder because the chiral spin texture is robust under the impurity scatterings\cite{Nomura2007}.
In fact, the chiral spin texture is observed in experiments\cite{Hsieh2009,Hsieh2009-2,Roushan2009}.
In the presence of disorder, the response function can be calculated by the replacement as
\begin{align*}
\mathcal{P}\frac{\omega/2}{\xi\pm\omega/2}\rightarrow\mathcal{P}\frac{\omega/2(\xi\pm\omega)}{(\xi\pm\omega/2)^2+1/\tau^2}
\end{align*}
with the relaxation time $\tau=2v/nu^2\mu_F$ in Eq. (\ref{int}).
The parameter $n$ and $u$ are the concentration and the strength of the impurities\cite{Suzuura2002}.
The impurity scatterings regularize the response function at $\omega=2\mu_F$ as 
\begin{align}
R_\mu(\omega)\simeq&\frac{\pi eg\mu_B\omega}{4}\ln
 (4\tau^2{\mu_F}^2+1).
\end{align}
\begin{figure}[htbp]
 \includegraphics[width=70mm]{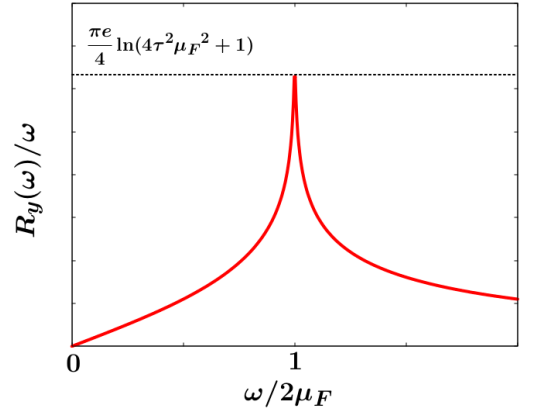}
\caption{The schematic picture of the dependence for the response function on the frequency.
A peak of the current is reached at $\omega=2\mu_F$.
 }\label{fig4}
\end{figure}
The imaginary part of the response function $\tilde{R}(\omega)$ can be written by
\begin{align*}
\tilde{R}_\mu(\omega)=&\frac{\pi}{eV}e^{\beta\Omega}\sum_{m,n} (e^{-\beta E_n}-e^{-\beta E_m})\\
&\times \langle n|j_\mu^S|m\rangle\langle m| j_\nu^S|n\rangle\delta(\omega+ \varepsilon_n -\varepsilon_m).
\end{align*}
The calculated results are expressed by,
\begin{align*}
\tilde{R}_\mu(\omega)=\frac{\pi^2\omega e}{2}\{f_F(\beta,-\omega/2-\mu_F)-f_F(\beta,\omega/2-\mu_F)\}.
\end{align*}
The imaginary part of the response function remains only when the initial state and intermediate state are occupied and empty, respectively. 

As discussed in Eq. (\ref{EffE}), when the chiral spin texture is present, the Zeeman term plays the same role as the temporally oscillating electric field in the surface states.
The effective electric field induces the electric current perpendicular to the magnetic field. 
Under the oscillating magnetic field, the electric current flows through only the surface because the Zeeman coupling does not induce the electric current in the bulk.
This is true even in the finite temperature or even in the presence of impurity scatterings.
We discuss why the current is absent in the bulk in the next section.
The surface electric current induced by the magnetic field can be detected separately from the electric current in the bulk of 3D TIs.

\section{Absence of Electromagnetic Induction in Bulk}
In the bulk of the TI, the electric states are described by $4\times4$ Hamiltonian,
\begin{align}
H=\varepsilon_0(p)I_{4\times4}+m(p)\beta_0+\boldsymbol{d}\cdot\boldsymbol{\alpha},\label{MDirac}
\end{align}
with
\begin{align*}
m(p)=&M_0-\{B_2({p_x}^2+{p_y}^2)+B_1{p_z}^2\},\\
\boldsymbol{d}(\boldsymbol{p})=&(A_2p_x,A_2p_y,A_1p_z),\;
\varepsilon_0(p)=C_0+C_1p^2,
\end{align*}
where $M_0$, $A_i$, $B_i$ and $C_i$ are the positive constants\cite{H.Zhang2009}.
The Dirac matrices $\alpha_\mu$ and $\beta_0$ are defined by
\begin{align*}
\alpha_\mu=\begin{pmatrix}
0&\sigma^\mu\\
\sigma^\mu&0
\end{pmatrix},\;
\beta_0=\begin{pmatrix}
\sigma^0&0\\
0&-\sigma^0
\end{pmatrix},
\end{align*}
with the $2\times2$ identity matrix $\sigma^0$.
In the following discussion, we ignore $\varepsilon_0(p)$ because it does not change the results.
The Hamiltonian so-called the massive Dirac Hamiltonian has the two degenerate eigenstates 
at a single momentum.
When a doubly degenerate eigenstate is represented by $\Psi_{\sigma}(p)$ with the spin eigenvalue $\sigma$,
the other eigenstate is
\begin{align*}
\mathcal{U}\Psi_{\sigma}(p)=\beta_0\Psi_{-\sigma}(p),
\end{align*}
because the massive Dirac Hamiltonian of Eq. (\ref{MDirac}) is invariant under the operation $\mathcal{U}$.
The degeneracy is preserved even in the presence of the non-magnetic disorder represented by the $4\times4$ unit matrix
because the operator $\mathcal{U}$ is independent of the momentum.
On the other hand, the Zeeman term changes its sign under the operation $\mathcal{U}$,
\begin{align}
\beta_0\{ \boldsymbol{B}\cdot(-\boldsymbol{\sigma})\}\beta_0^\dagger=-\boldsymbol{B}\cdot\boldsymbol{\sigma}.
\end{align}
The properties under the operation lead to the absence of the electric current under the temporally oscillating magnetic field.
In the bulk of 3D TIs, the electric current $\langle\boldsymbol{j}^B\rangle$ in the linear response theory is also represented by Eq. (\ref{KuboF}).
Since the density matrix in Eq. (\ref{KuboF}) represents the electric states in the absence of the Zeeman field,
the electric current is invariant under the exchange of $\rho_0\rightarrow \mathcal{U}^{-1}\rho_0\mathcal{U}$ as
\begin{align}
\langle j^B\rangle_\nu=\langle {j'}^B\rangle_\nu,\label{current1}
\end{align} 
with
\begin{align*}
\langle {j'}^B\rangle_\nu=-i\int^t_{-\infty}dt'\mathrm{Tr}\left\{\mathcal{U}^{-1}\rho_0\mathcal{U}\left[\hat{j}_\nu^S(t),\hat{H}'(t')\right]\right\}.
\end{align*}
Using the property of the trace and inserting $\mathcal{U}\mathcal{U}^{-1}=1$, the ${j'}_\nu^B$ is 
\begin{align*}
\langle {j'}^B\rangle_\nu=-i\int^t_{-\infty}dt'\mathrm{Tr}\left\{\rho_0\left[\mathcal{U}\hat{j}_\nu^S(t)\mathcal{U}^{-1},\mathcal{U}\hat{H}'(t')\mathcal{U}^{-1}\right]\right\}.
\end{align*}
Since the only $\hat{H}'(t')$ changes its sign under the operation $\mathcal{U}$,
the $\langle {j'}^B\rangle_\nu$ must satisfy
\begin{align}
\langle {j}^B\rangle_\nu=-\langle {j'}^B\rangle_\nu.\label{current2}
\end{align} 
As a consequence of Eqs. (\ref{current1}) and (\ref{current2}), the electric current under the temporally oscillating magnetic field is absent in the bulk of 3D TIs.
By applying the same argument, it is also possible to show the absence of the electric current in graphene. There are four Dirac cones in the pseudospin space on the two-dimensional Brillouine zone of graphene\cite{Neto2009,DasSarma2011}.
The two pairs of the Dirac cones are degenerate with respect to spin degree of freedom.
Therefore the electric transport phenomenon discussed in Sec. II is unique to the 3D TIs.

\section{Discussion}

\begin{figure}[htbp]
 \includegraphics[width=80mm]{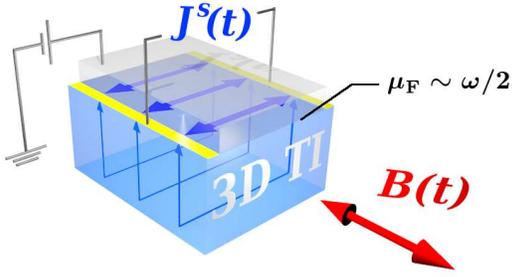}
\caption{The schematic picture of the realistic situation to measure the electric current.
The upper surface is insulating for the alternating current with the frequency $\omega$ by an bias voltage
whereas the lower surface still remains in the conductor.
 }\label{fig3}
\end{figure}

In the situation of Fig. \ref{fig1}, the alternating current induced by the temporally oscillating magnetic field flows on the all surfaces.
When we use the surface states in applications, it is necessary to separate a single surface as the transport device.
For this aim, we suggest a setup whose details are shown in Fig. \ref{fig3}.
The current peak can be observed on only the top surface when we tune the gate voltage so that the Fermi energy at the top surface to be half of the frequency.
In this case, since only the current through the top surface reaches a peak, the total current dominated by the top surface.

\section{Conclusion}
In conclusion, we have studied the novel transport phenomenon on the surface of three-dimensional topological insulators. 
The temporally oscillating magnetic field can induce the electric current on the surface because of the two characteristic features of the surface states: the linear dispersion and the chiral spin texture on the Fermi surface.
Owing to these features, the magnetic field plays the same role as the electric field.
In the bulk, on the other hand, the electric current is absent under the oscillating magnetic field because of the spin degeneracy in the bulk states.
Therefore the electric transport discussed in this paper is the purely surface effect.
We analytically calculate the response function of the electric current to the oscillating magnetic field.
In the clean limit at the zero-temperature, the electric current reaches a peak at $\omega=\mu_F$.
We also discuss how to detect the announced effect in experiments.

\section{acknowledgments}
This work was supported by 
the "Topological Quantum Phenomena" (Grant No. 22103002) Grant-in Aid for 
Scientific Research on Innovative Areas from the Ministry of Education, 
Culture, Sports, Science and Technology (MEXT) of Japan.

\appendix
\begin{widetext}
\section{Electric current in the linear response theory}
The homogeneous current density is represented by using the Kubo formula as
\begin{align*}
\langle j\rangle_\mu(t)=&-\frac{i}{V}\int_{-\infty}^t dt' \mathrm{Tr}\{\rho_0[\hat{j}_\mu^S(t),\hat{H}'(t')]\}\\
\hat{H}'(t)=&\hat{j}_\nu(t) \frac{eg\mu_BB(t)}{\omega}\\
\end{align*}
with $B(t)=B_x\exp[-i\omega t]$.
The current operator $j_\mu^S(t)$ is 
\begin{align*}
\hat{j}_\mu^S(t)=\sum_{\ell}\psi_{\ell}(t)j\psi^\dagger_{\ell}(t)
\end{align*}
with 
\begin{align*}
\psi^\dagger_{\ell}(t)=e^{iK_0t}\psi_{\ell}e^{-iK_0t},
\end{align*}
where $K_0=H_0-\mu_FN$ and $\psi_{\ell}$ are the total energy operator and the annihilation operator.
The commutator can be rewritten by 
\begin{align*}
[\hat{j}_\mu^S(t),\hat{H}'(t')]=\frac{eg\mu_BB(t)}{\omega}e^{i\omega(t-t')}[\hat{j}_\mu^S(t),\hat{j}_\nu^S(t')].
\end{align*}
Using the commutator, the current and external field relate to each other by the Fourier transformation,
\begin{align*}
\langle j\rangle_\mu(t)=&\frac{eg\mu_BB(t)}{V\omega}\int^{\infty}_{-\infty} dt'e^{i\omega t'} U_{\mathrm{ret}}(t'),
\end{align*}
with
\begin{align*}
U_{\mathrm{ret}}(t)=&-i\theta(t)\mathrm{Tr}\{\rho_0[\hat{j}_\mu^S(t),\hat{j}_\nu^S(0)]\},
\end{align*}
where $\theta(t)$ is the step function.
When the eigenstates $|n>$ for $K_0$ are known, the retarded function is represented by, 
\begin{align*}
U_{\mathrm{ret}}(t)=&-i\theta(t)\sum_n\langle n|\rho_0[\hat{j}_\mu^S(t),\hat{j}_\nu^S(0)]|n\rangle,\\
=&-i\theta(t)\sum_n e^{-\beta(E_n-\Omega)}\langle n|[\hat{j}_\mu^S(t),\hat{j}_\nu^S(0)]|n\rangle,
\end{align*}
where the $\Omega$ and $E_n$ are thermodynamic potential and the eigenvalue of $K_0$ for the state $|n\rangle$. 
Using the identity matrix $\sum_m|m\rangle\langle m|=1$,
\begin{align*}
U_{\mathrm{ret}}(t)
=&-i\theta(t)\sum_{m,n} e^{-\beta(E_n-\Omega)}\left(\langle n|\hat{j}_\mu^S(t)|m\rangle\langle m| \hat{j}_\nu^S(0)|n\rangle
-\langle n|\hat{j}_\nu^S(0)|m\rangle\langle m| \hat{j}_\mu^S(t)|n\rangle\right).
\end{align*}
The eigenstate $|n\rangle$ includes $N$ particles where one of the particles with $\varepsilon_n$,
\begin{align*}
U_{\mathrm{ret}}(t)
=&-i\theta(t)e^{\beta\Omega}\sum_{m,n} e^{-\beta E_n}\left(e^{-i(\varepsilon_m-\varepsilon_n)t}\langle n|\hat{j}_\mu^S|m\rangle\langle m| \hat{j}_\nu^S|n\rangle
-e^{i(\varepsilon_m-\varepsilon_n)t}\langle n|\hat{j}_\nu^S|m\rangle\langle m| \hat{j}_\mu^S|n\rangle\right).
\end{align*}
The total energy $E_n$ is equal to $E_0+\varepsilon_n$ where $E_0$ is the energy of $N-1$ particles. 
Exchanging $m$ and $n$ in the second term,
\begin{align*}
U_{\mathrm{ret}}(t)
=&-i\theta(t)e^{\beta\Omega}\sum_{m,n} (e^{-\beta E_n}-e^{-\beta E_m})\left(e^{-i(\varepsilon_m-\varepsilon_n)t}\langle n|\hat{j}_\mu^S|m\rangle\langle m| \hat{j}_\nu^S|n\rangle\right).
\end{align*}
The Fourier transform is 
\begin{align*}
U_{\mathrm{ret}}(\omega)
=&-i\int^{\infty}_{-\infty}dt e^{i(\omega+i\epsilon)t}\theta(t)e^{\beta\Omega}\sum_{m,n} (e^{-\beta E_n}-e^{-\beta E_m})\left(e^{-i(\varepsilon_m-\varepsilon_n)t}\langle n|\hat{j}_\mu^S|m\rangle\langle m|\hat{j}_\nu^S|n\rangle\right)\\
=&e^{\beta\Omega}\sum_{m,n} (e^{-\beta E_n}-e^{-\beta E_m}) \frac{\langle n|\hat{j}_\mu^S|m\rangle\langle m| \hat{j}_\nu^S|n\rangle}{\omega+ \varepsilon_n -\varepsilon_m+i\epsilon}.
\end{align*}
Thus the current is calculated as
\begin{align*}
\langle j\rangle_\mu(t)=&-\frac{{eg\mu_BB(t)}}{V\omega}e^{\beta\Omega}\sum_{m,n} (e^{-\beta E_n}-e^{-\beta E_m}) \langle n|\hat{j}_\mu^S|m\rangle\langle m|\hat{j}_\nu^S|n\rangle\frac{1}{\omega+ \varepsilon_n -\varepsilon_m+i\epsilon}\\
=&-\frac{{g\mu_BB(t)}}{V\omega}e^{\beta\Omega}\sum_{m,n} (e^{-\beta E_n}-e^{-\beta E_m}) \langle n|\hat{j}_\mu^S|m\rangle\langle m|\hat{j}_\nu^S|n\rangle\left(\mathcal{P}\frac{1}{\omega+ \varepsilon_n -\varepsilon_m}-i\pi\delta(\omega+ \varepsilon_n -\varepsilon_m)\right).
\end{align*}
We define a response function $\mathcal{R}_\mu=\langle j\rangle_\mu(t)/B(t)$ and separate the response function into the two terms as
\begin{align*}
R_\mu(\omega)=&-\frac{g\mu_B}{eV}e^{\beta\Omega}\sum_{m,n} (e^{-\beta E_n}-e^{-\beta E_m}) \langle n|\hat{j}_\mu^S|m\rangle\langle m| \hat{j}_\nu^S|n\rangle\mathcal{P}\frac{1}{\omega+ \varepsilon_n -\varepsilon_m}\\
\tilde{R}_\mu(\omega)=&\frac{\pi g\mu_B}{eV}e^{\beta\Omega}\sum_{m,n} (e^{-\beta E_n}-e^{-\beta E_m}) \langle n|\hat{j}_\mu^S|m\rangle\langle m|\hat{j}_\nu^S|n\rangle\delta(\omega+ \varepsilon_n -\varepsilon_m).
\end{align*}

It is easy to show that the real and the imaginary parts of the response function satisfy the Kramers-Kronig relation to each other.
In the case of $\mu=\nu$, the real part of the response function $R_\mu$ is
\begin{align*}
R_\mu(\omega)=&-\frac{g\mu_B}{eV}e^{\beta\Omega}\sum_{m,n} e^{-\beta E_n}|\langle n|\hat{j}_\mu^S|m\rangle|^2\left(\mathcal{P}\frac{1}{\omega+ \varepsilon_n -\varepsilon_m}-\mathcal{P}\frac{1}{\omega-(\varepsilon_n -\varepsilon_m)}\right)\\
\end{align*}
In the homogeneous system, the eigenstates $|m\rangle$ and $|n\rangle$ must have the same momentum $\boldsymbol{p}$.
In this case, the $R_\mu$ vanish when the one-particle states with $\varepsilon_m$ and $\varepsilon_n$ belong to the same cone.
Therefore, the current remains with non-zero value when the eigenstates are $|\boldsymbol{p},\pm\rangle$ and $|\boldsymbol{p},\mp\rangle$ respectively.
Since the difference $\varepsilon_n-\varepsilon_m$ is equal to $\pm 2vp$ with $p=|\boldsymbol{p}|$,
the response function is represented by
\begin{align*}
R_\mu(\omega)
=&-\frac{g\mu_B}{eV}e^{\beta\Omega}\sum_{m,n}e^{-\beta\Omega_{n}} |\langle\boldsymbol{p}_n,+|j_\mu^S|\boldsymbol{p}_n,-\rangle|^2V^{-1}\delta(\boldsymbol{p}_n-\boldsymbol{p}_m)
\left(\mathcal{P}\frac{1}{\omega\pm 2vp_n}-\mathcal{P}\frac{1}{\omega\mp 2vp_n}\right),
\end{align*}
with
\begin{align*}
e^{-\beta\Omega_n}=\langle e^{-\beta(E_0\pm2vp_n-\mu_FN)}\rangle_{N-1}
\end{align*}
where $\langle\cdots\rangle_{N-1}$ is an summation for states with any number of particles under the condition that one of the particles occupies the state with $|\boldsymbol{p}_n,\pm\rangle$.
The two exponent function of $e^{\beta\Omega}$ and $e^{\beta\Omega_n}$ can be calculated by the same manner as,
\begin{align*}
e^{-\beta\Omega}=\prod_{\boldsymbol{p}}\left(1+e^{-\beta(vp-\mu_F)}\right)\left(1+e^{-\beta(-vp-\mu_F)}\right)
\end{align*}
and
\begin{align*}
e^{-\beta\Omega_n}=&e^{-\beta(\varepsilon_n-\mu_F)}{\prod_{\boldsymbol{p}}}'\left(1+e^{-\beta(vp-\mu_F)}\right)\left(1+e^{-\beta(-vp-\mu_F)}\right)\\
=&\frac{e^{-\beta(\varepsilon_n-\mu_F)}}{1+e^{-\beta(\varepsilon_n-\mu_F)}}\prod_{\boldsymbol{p}}\left(1+e^{-\beta(vp-\mu_F)}\right)\left(1+e^{-\beta(-vp-\mu_F)}\right),
\end{align*}
where $\prod'$ is product of the all eigenstates without $n$.
Therefore, the product of $e^{-\beta(\Omega_n-\Omega)}$ is Fermi distribution function
\begin{align*}
e^{-\beta(\Omega_n-\Omega)}=&f_F(\beta,\varepsilon_n-\mu_F)\\
=&\frac{1}{1+e^{\beta(\varepsilon_n-\mu_F)}}.
\end{align*}
For the two-dimensional Dirac fermion, the spin of the eigenstate $|\boldsymbol{p},\pm\rangle$ depends only on the direction of the momentum $\hat{\boldsymbol{p}}$ and independent of $p$,
\begin{align*}
R_\mu(\omega)
=&-\frac{g\mu_B}{e}\int_0^{2\pi} d\theta_{\hat{\boldsymbol{p}}}|\langle \hat{\boldsymbol{p}},+|{j_\mu^S}|\hat{\boldsymbol{p}}-\rangle|^2\int^{p_g}_{0} dp\; p\left\{f_F(\beta,vp-\mu_F)
\left(\mathcal{P}\frac{1}{\omega+ 2vp}-\mathcal{P}\frac{1}{\omega- 2vp}\right)\right.\\
&\left.
+f_F(\beta,-vp-\mu_F)
\left(\mathcal{P}\frac{1}{\omega-2vp}-\mathcal{P}\frac{1}{\omega+2vp}\right)
\right\},
\end{align*}
where $2vp_g$ is the insulating gap in the bulk of the TI.

In the case of the low-temperature limit $\beta\rightarrow\infty$, the Fermi distribution function is equal to the step function,
\begin{align*}
\lim_{\beta\rightarrow\infty}f_F(\beta,x)=\theta(-x).
\end{align*}
The step function restricts the region of the integral in $[\mu_F,\;\xi_g]$ as
\begin{align}
R_\mu(\omega)
=&\frac{A}{2e}\int^{\xi_g}_{\mu_F} d\xi\;
\left(2-\mathcal{P}\frac{\omega/2}{\xi+\omega/2}+\mathcal{P}\frac{\omega/2}{\xi-\omega/2}\right),\label{apint}
\end{align}
with 
\begin{align*}
A=g\mu_B\int_0^{2\pi} d\theta_{\hat{\boldsymbol{p}}}|\langle \hat{\boldsymbol{p}},+|\frac{j_\mu^S}{v}|\hat{\boldsymbol{p}}-\rangle|^2
\end{align*}
When the half of frequency is smaller than the Fermi energy $\omega/2<\mu_F$, there is no singularity in the region of integral as
\begin{align*}
R_\mu(\omega)
=&\frac{A}{2e}
\left\{2(\xi_g-\mu_F)-\frac{\omega}{2}\ln\frac{\xi_g+\omega/2}{\mu_F+\omega/2}+\frac{\omega}{2}\ln\frac{\xi_g-\omega/2}{\mu_F-\omega/2}\right\}\\
=&\frac{A}{2e}
\left\{2(\xi_g-\mu_F)+\frac{\omega}{2}\ln\frac{\xi_g-\omega/2}{\xi_g+\omega/2}+\frac{\omega}{2}\ln\frac{\mu_F+\omega/2}{\mu_F-\omega/2}\right\}
.
\end{align*}
On the other hand, in the case of $\mu_F<\omega/2$, there is a singular point at $\xi=\omega/2$,
\begin{align*}
R_\mu(\omega)
=&\frac{A}{2e}
\left\{2(\xi_g-\mu_F)-\frac{\omega}{2}\ln\frac{\xi_g+\omega/2}{\mu_F+\omega/2}
+\frac{\omega}{2}\left(\int^{\xi_g}_{\omega+\delta} d\xi+\int_{\mu_F}^{\omega-\delta}d\xi\right)
\;\frac{1}{\xi-\omega/2}\right\}\\
=&\frac{A}{2e}
\left\{2(\xi_g-\mu_F)-\frac{\omega}{2}\ln\frac{\xi_g+\omega/2}{\mu_F+\omega/2}
+\frac{\omega}{2}\lim_{\delta\rightarrow0}\left(\int^{\xi_g}_{\omega/2+\delta} \frac{d\xi}{\xi-\omega/2}+\int_{\omega/2-\mu_F}^{\delta}
\;\frac{d\xi'}{\xi'}\right)\right\}\\
=&\frac{A}{2e}
\left\{2(\xi_g-\mu_F)-\frac{\omega}{2}\ln\frac{\xi_g+\omega/2}{\mu_F+\omega/2}
+\frac{\omega}{2}\lim_{\delta\rightarrow0}\left(\ln \frac{\xi_g-\omega/2}{\delta}+\ln
\;\frac{\delta}{\omega/2-\mu_F}\right)\right\}\\
=&\frac{A}{2e}
\left\{2(\xi_g-\mu_F)-\frac{\omega}{2}\ln\frac{\xi_g+\omega/2}{\mu_F+\omega/2}
+\frac{\omega}{2}\ln \frac{\xi_g-\omega/2}{\omega/2-\mu_F}\right\}\\
=&\frac{A}{2e}
\left\{2(\xi_g-\mu_F)+\frac{\omega}{2}\ln\frac{\xi_g-\omega/2}{\xi_g+\omega/2}
+\frac{\omega}{2}\ln \frac{\mu_F+\omega/2}{\omega/2-\mu_F}\right\}
.
\end{align*}
Therefore, the current reaches a peak at the vicinity of $\mu_F=\omega/2$ as
\begin{align}
R_\mu(\omega)\simeq&\frac{\omega A}{4e}\ln\left|\frac{2\mu_F}{\mu_F-\omega/2}\right|\label{peak}.
\end{align}

\end{widetext}
\bibliography{TI}

\end{document}